\documentclass{aa}  

\usepackage{subcaption}
\usepackage{natbib}
\bibpunct{(}{)}{;}{a}{}{,} 
\usepackage{graphicx}
\usepackage{multirow}
\usepackage{amsmath}
\usepackage{newtxtext,newtxmath}
\usepackage{lscape}
\usepackage{subcaption}
\usepackage{gensymb}

\usepackage{tabularx}
\usepackage{graphicx}
\usepackage{lipsum}

\usepackage{placeins} 
\usepackage{float}

\usepackage[separate-uncertainty=true,multi-part-units=single]{siunitx}
\DeclareSIUnit[]{\angstrom}{\textup{\AA}}
\DeclareSIUnit[]{\Rsun}{\text{\ensuremath{R_{\sun}}}}
\DeclareSIUnit[]{\Msun}{\text{\ensuremath{M_{\sun}}}}
\DeclareSIUnit[]{\kms}{km/s}

\usepackage{pgfplots}
\usepackage{color}
\usepackage[colorinlistoftodos,prependcaption,textsize=tiny,color=red]{todonotes}

\usepackage[]{hyperref}
\hypersetup{
  colorlinks   = true, 
  urlcolor     = blue, 
  linkcolor    = blue, 
  citecolor   = blue 
}
\begin{document} 
   \title{Eclipsing white dwarf from the Zwicky Transient Facility}
   \subtitle{II. Seven eclipsing double white dwarfs}

   \author{J. van Roestel\thanks{corresponding author: jcjvanroestel.astro@gmail.com} \inst{\ref{uva}} \and \\ 
    K.~Burdge \inst{\ref{mit}} \and 
    I.~Caiazzo \inst{\ref{ista}} \and 
    T.~Kupfer \inst{\ref{hamburg}} \and 
    P.~Mr\'oz \inst{\ref{warszaw}} \and 
    T.A.~Prince \inst{\ref{caltech}} \and 
    A.C.~Rodriguez \inst{\ref{caltech}} \and 
    S.~Toonen\inst{\ref{uva}} \and \\
    Z.~Vanderbosch  \inst{\ref{caltech}} \and \\
    E.C.~Bellm\inst{\ref{UW}} \and 
    A.J.~Drake\inst{\ref{caltech}} \and 
    M.J.~Graham\inst{\ref{caltech}} \and
    S.L.~Groom\inst{\ref{ipac}} \and
    G.~Helou\inst{\ref{ipac}} \and
    S.R.~Kulkarni\inst{\ref{caltech}} \and 
    A.A.~Mahabal  \inst{\ref{caltech}} \and  
    R.L.~Riddle\inst{\ref{coo}} \and
    B.~Rushome\inst{\ref{ipac}}  
        }
        
   \institute{Anton Pannekoek Institute for Astronomy, University of Amsterdam, 1090 GE Amsterdam, The Netherlands\label{uva}
   \and Department of Physics, Massachusetts Institute of Technology, Cambridge, MA 02139, USA\label{mit}
   \and Institute of Science and Technology Austria, Am Campus 1, 3400 Klosterneuburg, Austria\label{ista}
   \and Hamburger Sternwarte, University of Hamburg, Gojenbergsweg 112, 21029 Hamburg, Germany\label{hamburg}
   \and Astronomical Observatory, University of Warsaw, Al. Ujazdowskie 4, 00-478 Warszawa, Poland\label{warszaw}
   \and Cahill Center for Astrophysics, California Institute of Technology, Pasadena, CA 91125, USA\label{caltech}
    \and DIRAC Institute, Department of Astronomy, University of Washington, 3910 15th Avenue NE, Seattle, WA 98195, USA\label{UW}
   \and IPAC, California Institute of Technology, 1200 E. California Blvd, Pasadena, CA 91125, USA\label{ipac}
   \and Caltech Optical Observatories, California Institute of Technology, Pasadena, CA 91125, USA\label{coo}
        }

   \date{Received May 21, 2025; accepted XX XX, XXXX}

\titlerunning{Seven eclipsing double white dwarfs from ZTF}

 
  \abstract
    {In a systematic search for eclipsing white dwarfs using Zwicky transient facility (ZTF) data, we found seven eclipsing double white dwarfs with orbital periods ranging from 45 minutes to 3 hours. We collected high-speed light curves, archival multi-wavelength data, and optical spectra for all systems and determined the binary parameters for each of them. We show that six of the systems are low-mass, double helium-core white dwarf binaries, with the last one a carbon-oxygen -- helium core white dwarf binary. These binaries slowly spiral inwards due to gravitational wave energy losses and are expected to merge within \qty{36}{Myr}--\qty{1.2}{Gyr}, and we predict that the shortest orbital period binary will show a measurable eclipse arrival time delay within a decade. The two longest systems show a delay in the arrival time of the secondary eclipse, which we attribute to a small eccentricity of $\approx 2\times10^{-3}$. This is the first time that a non-zero eccentricity is measured in a compact double white dwarf binary. We suggest that these systems emerged from the common envelope with this small eccentricity, and because of the relatively long orbital period, gravitational wave emission has not yet circularised the binaries. Finally, we predict that relativistic apsidal precession will result in a change in the delay of the secondary eclipse that is measurable within a decade.
    } 
   \maketitle


%


\section{Introduction}\label{sec:intro}
Compact double white dwarfs (dWD) are binary systems composed of two white dwarf stars with orbital periods that range from several days to as short as a few minutes. They are of astrophysical interest because they are descendants of multiple mass-transfer phases.
Compact dWD binaries are compact enough that they are strong gravitational wave sources and may be directly observable with upcoming space-based gravitational wave detectors like TianQuin \citep{luo2016} and the Laser Interferometer Space Antenna \citep[LISA,][]{amaro-seoane2017,amaro-seoane2023}. Double white dwarf binaries are the dominant Galactic gravitational wave sources \citep{nissanke2012,korol2017,lamberts2019}, and binaries have already been found that will be detected by LISA (see \citealt{kupfer2024} for a recent overview). The dWD binary systems with orbital periods shorter than \qty{10}{h} will come into contact within a Hubble time, and will start stable mass-transfer or merge into a single object \citep{shen2015}. The final outcome depends on the mass, and they are the supposed progenitors of exotic merger stars \citep{webbink1984,zhang2012,cheng2019,gvaramadze2019,hollands2020} and are the likely progenitors of Type Ia supernovae \citep{maoz2014,munday2025}.

 The Galactic population of dWD binaries is expected to be in the hundreds of millions \citep{nelemans2001a,marsh2011,korol2022}, but because they are faint, only $\approx$300 double white dwarf binaries\footnote{\url{https://github.com/JamesMunday98/CloseDWDbinaries}, version 20241121} have been identified so far. The majority of these systems have been found by spectroscopic searches: SN Ia Progenitor surveY \citep[SPY][]{napiwotzki2020}, Extremely Low Mass (ELM) survey \citep{brown2020,kosakowski2023}, and most recently double-lined double white dwarf (DBL) survey \citep{munday2024}. These surveys are responsible for the discovery of most of the systems, and are useful to understand the population statistics. For example, \citet{napiwotzki2020} measured a dWD fraction of 6\% compared to the general sample of white dwarfs, which is consistent with the fraction expected from population synthesis models \citep{toonen2017}. 

For detailed characterisation of individual systems, eclipsing dWDs are extremely useful. First, eclipsing systems can be used to precisely measure model-independent masses and radii of white dwarfs \citep{parsons2017a}. Double white dwarfs are also useful to test the common envelope efficiency; how efficient orbital energy is used to unbind the envelope \citep[e.g.][]{nelemans2001a,vandersluys2006,scherbak2023}. If one of the white dwarfs is pulsating, they can be used to empirically constrain the core composition of low-mass stellar remnants and to investigate the effects of close binary evolution on the internal structure of white dwarfs \citep{parsons2020}.
The precise timing of the eclipses can also be used to test a variety of physics. The emission of gravitational waves results in the slow decay of the orbit \citep{hermes2014,burdge2019}. Eclipse timing can also measure the orbital decay due to tides \citet{carvalho2022}.
Double-eclipsing white dwarfs (where both the primary and secondary eclipse is detected) can be used to measure the eccentricity \citep{kaplan2010,kaplan2014}, although no double white dwarf with a non-zero eccentricity has been found so far. Finally, eclipsing timing campaigns of eclipsing double white dwarfs can be used to reveal the presence of third objects \citep[e.g.][]{marsh2014}. Crucially, double white dwarf binaries are not expected to suffer from orbital period changes due to Applegate mechanism \citep{applegate1992} as opposed to eclipsing white dwarfs with red dwarf companions \citep{bours2016}, making them very stable `clocks' \citep{shaw2025} .

A total of 19 eclipsing double white dwarfs are currently known. The first two discovered are NLTT~11748 \citep{steinfadt2010} and GALEX~J1717+6757 \citep{vennes2011}, both with orbital periods of about 6 hours. These were quickly followed by the discovery of CSS~41177 \citep{parsons2011} with a period of 3 hours. SDSS~J0651+2844 was the fourth discovered eclipsing double white dwarf \citep{brown2011} and has a much shorter orbital period of only 12~minutes. In the following decade, more systems were discovered: SDSS~J0751\textminus0141 \citep{kilic2014}, SDSS~J1152+0248 \citep{hallakoun2016}, SDSS~J0822+3048 \cite{brown2017}. 

The discovery rate increased dramatically when the Zwicky Transient Facility (ZTF) started operating in 2018 \citep{graham2019,bellm2019,masci2019,dekany2020}. Using ZTF, \citet{burdge2020} systematically searched for short orbital period (<\qty{1}{hr}) white dwarf binaries and discovered seven eclipsing double white dwarfs. This includes the discovery of a 7~minute orbital period double eclipsing white dwarf, the shortest orbital period yet for an eclipsing double white dwarf \citep{burdge2019a}. As ZTF continued to observe, more systems were discovered \citep{coughlin2021,keller2021}. In addition, more eclipsing double white dwarfs are still discovered by other surveys, for example  WD~J0225\textminus6920 \citep{munday2023}, WD~J0221+1710 \citep{kosakowski2023} ELM survey, WD~J2102\textminus4145 \citep{antunesamaral2024}.

In this paper, we present the discovery (or re-discovery) and characterisation of seven eclipsing double white dwarf systems found in a systematic search of the ZTF data for periodically eclipsing white dwarfs with periods of $>$\qty{60}{minutes} 
(Van Roestel et al. 2025, in prep)\footnote{the 45 minute period system was initial detected at twice the orbital period of 90 minutes}. We note that one system was also identified by the ELM survey \citet{kosakowski2023}, and two systems were identified as eclipsing by \citet{keller2021} using ZTF data.
For all systems, we obtained high-speed followup photometry and identification spectra, and we obtained phase resolved spectra for four systems and ID-spectra for the other three, see Section~\ref{sec:data}. In Section~\ref{sec:analysis} we show the analysis of data and Section~\ref{sec:results} presents the binary parameters for all seven systems. In Section~\ref{sec:discussion}, we discuss how these binaries compare to other eclipsing double white dwarfs, the first detection of eccentric double white dwarfs, and the in-spiral due to gravitational wave emission and the final fate of these systems. We discuss future work in Section~\ref{sec:future} and summarise the paper in Section~\ref{sec:conclusion}.

\begin{table*}
\centering
\caption{Overview of the seven eclipsing double white dwarfs stars discovered by their eclipses in the ZTF light curves. The parallax is taken from \textit{Gaia} eDR3 \citet{brown2020} and the distance is from \citet{bailer-jones2021}. The dust-extinction is taken from \citet{green2019a}. $\mathrm{^1}$: also found by \citet{kosakowski2023} in the ELM-South survey and $\mathrm{^2}$ were also identified as eclipsing double white dwarfs by \citet{keller2021} using ZTF data.} \label{tab:WDoverview}
\renewcommand{\arraystretch}{1.25}
\begin{tabular}{lllllllllll}
Name & RA & Dec & $P_\mathrm{orb}$ &  & $G$ & $BP-RP$ & parallax & distance \\
 & hh:mm:ss.s & dd:mm:ss.s  & hours &  &  Vega-mag & Vega-mag & mas & pc \\
\hline\hline
ZTF J0221+1710$^1$ & 02:21:10.8 &  \phantom{+}17:10:49.1 & 1.47 & & 17.68 & \phantom{\textminus}0.13 & 3.58 $\pm$ 0.12 & $271^{+10}_{-7}$ \\
ZTF J0238+0933 & 02:38:35.0 &  \phantom{+}09:33:03.5 & 3.28 & & 18.64 & \textminus0.08 & 1.11 $\pm$ 0.23 & $921^{+159}_{-151}$ \\
ZTF J0720+6439 & 07:20:03.0 &  \phantom{+}64:39:47.4 & 0.75 & &19.01 & \textminus0.11 & 1.14 $\pm$ 0.23 & $921^{+256}_{-161}$ \\
ZTF J1110+7445 & 11:10:16.7 &  \phantom{+}74:45:59.9 & 2.89 & & 18.62 & \phantom{\textminus}0.00 &	1.97 $\pm$ 0.16 & $514^{+47}_{-44}$ \\
ZTF J1356+5705$^2$ & 13:56:26.7 &  \phantom{+}57:05:46.0 & 1.53 & & 18.94 & \phantom{\textminus}0.33 &	2.80 $\pm$ 0.18 & $361^{+27}_{-29}$ \\
ZTF J1758+7642$^2$ & 17:58:12.9 & \phantom{+}76:42:16.9 & 3.15 & & 18.96 & \textminus0.22 &	1.61 $\pm$ 0.19 & $624^{+69}_{-54}$ \\
ZTF J2249+0117 & 22:49:01.6 & \phantom{+}01:17:22.7 & 2.30 & & 18.55 & \textminus0.14 & 1.62 $\pm$ 0.20 & $647^{+124}_{-74}$ \\
\hline
\end{tabular}
\end{table*}

\section{Data}\label{sec:data}

\begin{figure*}
    \includegraphics{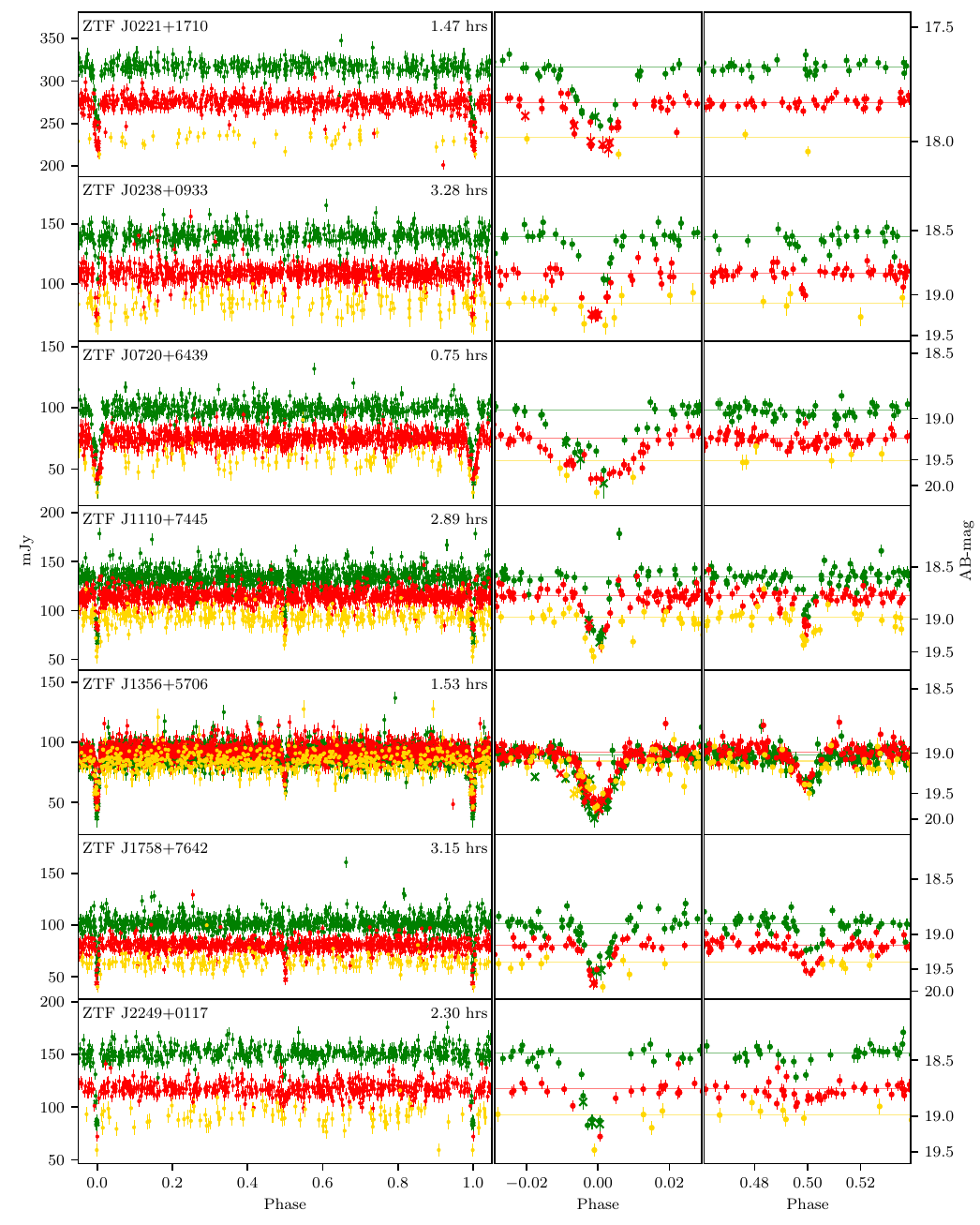}
    \caption{The ZTF lightcurves of the double white dwarfs folded to their orbital periods (shown in the top right). The panels on the right show the primary and secondary eclipses. Data from the $g$, $r$, and $i$-bands are displayed in green, red, and gold. Dots show PSF-photometry, while crosses show image-differencing (`alert') photometry.}
    \label{fig:ZTFlcs}
\end{figure*}

\subsection{Zwicky Transient Facility light curves}

\begin{figure}
    \centering
    \includegraphics[]{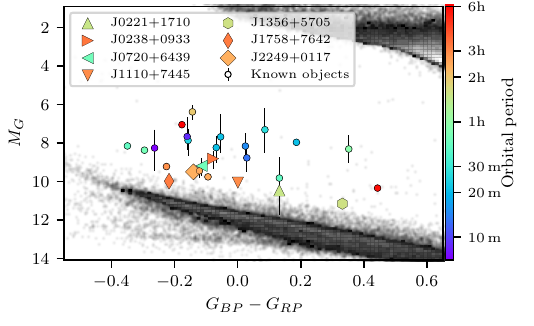}
    \caption{Colour-magnitude diagram of all known eclipsing double white dwarfs. The colours indicate the orbital period. The magnitude and distance are based on \textit{Gaia} dr3 data \citep{brown2020b,bailer-jones2021}.}
    \label{fig:HRdiagram}
\end{figure}

As part of the Zwicky Transient Facility (ZTF), the Palomar 48-inch (P48) telescope images the sky every night. 
Most observations use the $g$ and $r$ filters, but a small fraction of the observations are done using the $i$ filter. The exposure times are predominantly 30 seconds for $g$ and $r$ and 60 seconds for $i$. The limiting magnitude is $\approx 20.5$ in all three bands. ZTF images are automatically processed, and two main data products are generated. The first are the `alerts' which are based on difference imaging and are mainly designed to identify transients. The second main data product is PSF-photometry of persistent sources in the science images \citep[for a full description see ][]{masci2019}.

The objects presented in this paper were initially identified as eclipsing white dwarfs in the search for eclipsing white dwarfs with period of more than 1 hour, see Van Roestel et al. (2025, in prep.). 
In short, we searched the ZTF-lightcurves of all sources in the \textit{Gaia} eDR3 white dwarf catalogue \citep{gentilefusillo2021} for periodic eclipses using the Box Least Squares method \footnote{\url{https://github.com/johnh2o2/cuvarbase}} \citep{kovacs2002}. Several systems were identified as candidate dWD binaries because of the lack of any infrared excess  from a non-degenerate companion or the possible detection of the secondary eclipse in the ZTF data. These systems were followed up with spectra and high speed photometry. This resulted in the discovery of seven new eclipsing double white dwarfs that were not known before the start of ZTF. Figure~\ref{fig:ZTFlcs} shows the ZTF lightcurves of the seven long-period eclipsing double white dwarf systems we discovered and Fig. \ref{fig:HRdiagram} shows their location in the \textit{Gaia} colour-magnitude diagram.  

We note that ZTF~J0720+6439 has a period of 45 minutes but was initial detected at twice the orbital period of 90 minutes. Although technically it falls outside the search parameters, the system is new and therefore we include it in this paper anyway. In addition, we also recovered two (CSS~41177 and SDSS~J1152+0248) of the four known eclipsing dWDs that have a period of more than 1 hour and are in the ZTF footprint.
The other two systems were not recovered because the eclipses are too shallow. We also identified a few false positives, white dwarf--brown dwarf systems that show grazing eclipses, and these will be presented in a different paper.

\subsection{Archival photometry}
For each object in the sample, we collect multi-wavelength data obtained from multiple surveys: UV data from \textit{GALEX} \citep{bianchi2017} and optical data from Pan-STARRS \citep{chambers2016}, SDSS \citep{gunn2006}, and \textit{Gaia} eDR3 \citep{gaiacollaboration2021}. In addition, we also use the parallax from \textit{Gaia}. 
For each object, we checked if there was no nearby object that could affect the photometry. We used \textit{Vizier} and \textit{astroquery} to cross-match and collect the data.

\subsection{Followup observations}
We obtained followup high-speed light curves, identification spectra, and phase resolved spectra of a few of the targets. An overview of all followup observations is given in Table~\ref{tab:followup}. We briefly describe the data collection and processing in this section.

\subsubsection{CHIMERA fast cadence photometry}
The Caltech HIgh-speed Multi-color camERA \citep[CHIMERA][]{harding2016} is a dual-channel photometer which uses frame-transfer, electron-multiplying CCDs mounted on the Hale 200-inch (5.1 m) Telescope at Palomar Observatory (CA, USA). The pixelscale is 0.28 arcsec/pixel (unbinned). We used the conventional amplifier, and used 2x2 binning on most nights (except when to seeing was excellent) to reduce the readout noise and readout time. Each of the images were bias subtracted and divided by twilight flat fields\footnote{\url{https://github.com/caltech-chimera/PyChimera}}.

Table~\ref{tab:CHIMERA} gives an overview of all CHIMERA observations. We obtained CHIMERA data for each target in the $g$ and $r$ filters and, in some cases, the $i$ filter. 
The seeing was typically between 0.8 and 1.5 arcsec. CHIMERA observations were typically carried out during grey or dark time.

We used the ULTRACAM pipeline \citep{dhillon2007} to perform aperture photometry. We used an optimal extraction method with a variable aperture of 1.5 the full-width-half-max (FWHM) of the seeing as measured from a nearby reference star. A differential lightcurve was created by simply dividing the counts of the target by the counts from the reference star. Timestamps of the images were determined using a GPS receiver.

\subsubsection{LRIS spectroscopy}

\begin{figure}
    \centering
    \includegraphics[]{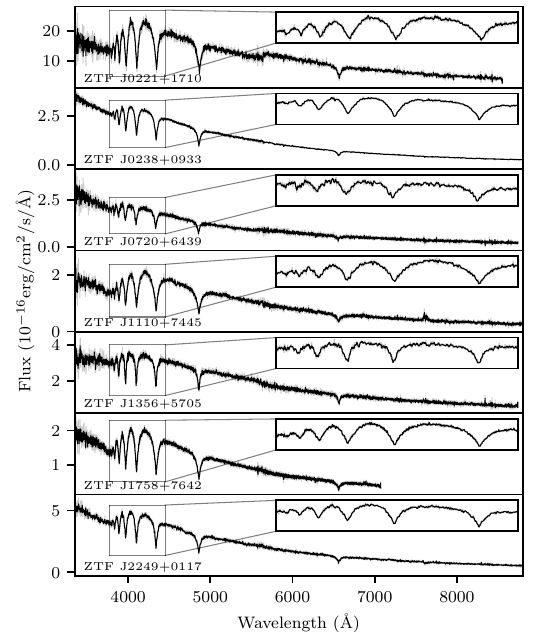}
    \caption{The LRIS spectra of each of the systems. All spectra show DA spectral type; no lines other than Hydrogen lines are detected.}
    \label{fig:spectra}
\end{figure}

Spectra of all seven systems was obtained using the Keck\,I Telescope (HI, USA) and the Low Resolution Imaging Spectrometer (LRIS; \citealt{Oke1995,McCarthy1998}). We used a $1^{\prime\prime}$ slit for all observations with various setups. For the blue arm, we used the R600 grism that has a resolution of $R\approx1100$. In the red arm, we used three different gratings, R400/8500, R600/7500, and R1200/7500. These setups result in resolutions of $R\approx950$, $R\approx1400$, and $R\approx2800$.

A standard longslit data reduction procedure was performed with the \textsc{Lpipe} pipeline\footnote{http://www.astro.caltech.edu/\texttildelow dperley/programs/lpipe.html} \citep{perley2019}. The pipeline reduced LRIS spectral data to spectra using the standard procedure, including calibration with a standard star. Wavelength calibration was done using lamp-spectra obtained at the beginning or end of each sequence. Fine-tuning of the wavelength calibration was done using sky-lines (which is standard procedure with \textsc{Lpipe}). 

\subsubsection{ESI spectroscopy}
We obtained phase resolved spectra of ZTF~J0221+1710 using the Keck\,II Telescope (HI, USA) and the Echellette Spectrograph and Imager (ESI; \citealt{sheinis2002}).
We used a $0.75^{\prime\prime}$ slit and 2x1 binning (spatial x spectral). The typical resolution with this setup is $R\approx$ \qty{26000}{}, with a spectral range from \qty{3800}{\angstrom} to \qty{10000}{\angstrom}. The data was reduced with the \textsc{Makee} pipeline\footnote{https://sites.astro.caltech.edu/\texttildelow tb/makee/} which performs a standard extraction procedure. We did not flux calibrate the spectra.

\section{Analysis}\label{sec:analysis}

\subsection{Spectral modelling and radial velocity amplitudes}

\begin{table}[]
    \centering
    \begin{tabular}{l|ll}
        Name & $K_1$ (km/s) & $K_2$ (km/s)\\
        \hline
        ZTF~J0221+1710 & $360\pm4$ & - \\
        ZTF~J0238+0933 & $166\pm16$ & $183\pm34$ \\
        ZTF~J0720+6439 & $272\pm15$ & $328\pm60$ \\
        ZTF~J1356+5706 & $166\pm8$ & $296\pm10$ \\
    \end{tabular}
    \caption{Radial velocity amplitudes measured from phase resolved spectra.}
    \label{tab:RVtable}
\end{table}

We obtained spectra for all objects with LRIS. These spectra show that all objects are DA type, only Balmer lines are detected, see Fig.~\ref{fig:spectra}. The quality and number of spectra per object vary. For four objects: ZTF~J0221+1710, ZTF~J0238+0933, ZTF~J0720+6439, ZTF~J1356+5705, we have sufficient spectra and were able to measure the radial velocity semi-amplitude. To do this, we model the spectral lines with DA model spectra. We used a slightly different approach for each object because of differences in the amount and quality of the data and relatively luminosity contributions from the two white dwarfs. The measured velocities are listed in Table~\ref{tab:RVtable} and the detailed analysis are discussed in the rest of this Section. 

We were unable to measure any radial velocity amplitude for the three other binaries: ZTF~J1110+7445, ZTF~J1758+7642, and ZTF~J2249+0117. 
The spectra of ZTF~J1110+7445 are dominated by the primary white dwarf and can be modelled with a single white dwarf spectrum. However, the signal-to-noise ratio of the data we obtained was not good enough to meaningfully constrain the velocity.
The eclipse depths of ZTF~J1758+7642 suggest that this is a double-lined system. We attempted to fit the data with two components moving in anti-phase, but the signal-to-noise ratio is simply too low to get any meaningful results.
ZTF~J2249+0117 was discovered relatively recently, and only a single identification spectrum was obtained. Based on the eclipse depths, the primary white dwarf contributes 80\% and is likely a single-line binary. 

\subsubsection{ZTF~J0221+1710}
This is the brightest source in our sample, and we got phase-resolved high-resolution ESI spectra. The secondary eclipse in this system is very shallow and barely detected, which suggests that the secondary star contributes only a small amount to the total luminosity ($\lesssim 5\%$ from the detailed binary modelling). We therefore simply modelled the spectra with just a single DA white dwarf model. We model each individual line and multiple the model spectrum with a third-order polynomial to account for the instrumental response (the ESI spectra are not flux-calibrated). For each of the 11 spectra, we fitted the H$\alpha$, H$\beta$, H$\gamma$, and H$\delta$ lines and determined the radial velocity shift of each, resulting in 44 velocity measurements. 

We then fitted a sinusoidal radial velocity curve to the velocity measurements using the period and mid-eclipse time from the light curve. For the ESI data, a closer inspection shows that there are systemic velocity offsets of $\approx$ \qty{20}{km/s} between the different lines. Therefore, we model the velocities with an independent systemic velocity shift for each line but with a shared velocity amplitude. The final velocity amplitude measured is $K_1$=\qty{360 \pm 4}{km/s}, see Fig~\ref{fig:RVcurve_ZTFJ0221_ESI}. 

We also fitted the LRIS data. For the LRIS data, we fit the blue spectra from \qty{3800}{\angstrom} to \qty{5000}{\angstrom} using the DA spectra. We again added a third-order polynomial to the model to account for any trends in the data (e.g. slit losses). The radial velocities from the LRIS data are $K_1$=\qty{336 \pm 11}{km/s}.
This is a 2 standard deviation lower compared to the ESI data. We also note that the radial velocity amplitude for this object was already measured by \citet{kosakowski2023}, $K_1$=\qty{347.9 \pm 4.2}{km \per s}, slightly different from what we determined with ESI. In the rest of this paper, we use the ESI-measured radial velocity amplitude.

\subsubsection{ZTF~J0238+0933}
The light curve modelling of this system suggests that the luminosity ratio is about 70\% to 30\%. Therefore, one white dwarf likely dominates the spectra, but the other cannot be ignored. For this object, we obtained one orbit of high-resolution red-arm LRIS spectra with a resolution of $R\approx 2600$. A visual inspection of this data shows a resolved line core moving on the orbital period.

We modelled these spectra with two DA model spectra. The model parameters for each individual star are: ($T_\mathrm{1,2}$), surface gravity ($\log g_\mathrm{1,2}$), radius ($R_\mathrm{1,2}$), velocity amplitude ($K_\mathrm{1,2}$) and systemic velocity ($\gamma_\mathrm{1,2}$). The binary parameters are distance ($D$) and reddening ($E_\mathrm{gr}$), and we use a fixed value for the orbital period and mid-eclipse time. Finally, for each individual spectrum, we use a separate scaling parameter and a parameter to add additional noise. We fit the spectra around the H$\alpha$ line, see Fig.~\ref{fig:spectrafit_ZTFJ0238}. We use emcee to find the best solution and uncertainties on all parameters (a total of 34 parameters).

Although the absorption lines are relatively subtle, by fitting all 11 spectra at the same time, we were able to constrain both the primary and secondary radial velocity amplitudes: $K_1=$\qty{166\pm16}{km/s} and $K_2=$\qty{183\pm34}{km/s}. 
By just fitting the H$\alpha$ lines, most parameters besides the radial velocity amplitudes are poorly constrained and are not useful.

\subsubsection{ZTF~J0720+6439}
For ZTF~J0720+6439, we collected 11 consecutive spectra with an exposure time of \qty{240}{s} spanning from phase 0.1 to 0.9. The second star contributes about 11\% around H$\alpha$, less in at short wavelengths (5\%). Even though the secondary white dwarf contributes only a little to the overall spectrum, we do fit the spectra with a model that consists of two white dwarfs moving in anti-phase (see above). In this case, we fit the blue LRIS spectra from \qty{3800}{\angstrom} to \qty{5100}{\angstrom}. As can be seen in Figure~\ref{fig:spectrafit_ZTFJ0720}, the deepest absorption lines clearly moves, but the secondary white dwarf is not clearly visible in a single spectrum. However, we do detect the motion of both white dwarfs, with $K_1=$ \qty{272\pm15}{km/s} and $K_2=$ \qty{328\pm60}{km/s}. 



\subsubsection{ZTF~J1356+5705}
For this object, we obtained 23 spectra with LRIS spanning $\approx$1.5 orbit. The primary and secondary eclipse depth suggests that both stars are roughly equally luminous (about 60\%-40\%). A visual inspection of the spectra clearly shows two lines moving in anti-phase with each other on the orbital period of the system. We therefore we the same method as for ZTF~J0238+0933, were we fit all spectra at the same time with a fixed period and phase; a model with 58 free parameters, see Fig.~\ref{fig:spectrafit_ZTFJ1356}. We limit the fit to the spectral range of \qty{3800}{\angstrom} to \qty{5100}{\angstrom}. The measured radial velocity amplitudes are $K_1=$\qty{166\pm8}{km/s} and $K_2=$\qty{296\pm10}{km/s}.

\subsection{Binary star modelling}
\subsubsection{The binary star model}
To determine the binary parameters for each system, we jointly modelled the combined light curves (both ZTF light curves and CHIMERA light curves), archival multi-wavelength photometry, and \textit{Gaia} parallax. To calculate the absolute magnitude in each filter we interpolate the pre-calculated magnitudes\footnote{\url{https://www.astro.umontreal.ca/~bergeron/CoolingModels/}} \citep{holberg2006,blouin2018,bedard2020}. To account for dust extinction, we use the reddening prescription by \citep{fitzpatrick1999} as implemented by the \textsc{extinction} python package. To generate the model light curves we use \textsc{ellc} \citep{maxted2016}. To calculate the surface brightness ratio in each filter we use the same pre-calculated magnitudes as above. 

The binary parameters of the model are the orbital period ($P$), mid-eclipse time of the primary eclipse ($t_0$), inclination ($i$), distance ($D$), and reddening ($E_{gr}$). For each of the stars, we have a mass ($M_\mathrm{1,2}$), radius ($R_\mathrm{1,2}$), temperature ($T_\mathrm{1,2}$). Because the stars are well within the Roche-lobe, we use a spherical model for both stars. In addition, we use fixed linear limb- and gravity-darkening parameters. For convenience, we also calculate a few other parameters which are: the semi-major axis ($a$), the mass-ratio $q=\frac{M_2}{M_1}$, the surface gravity of both stars $\log g_\mathrm{1,2}$ (in cgs~units), $K_\mathrm{1,2}$ are the radial velocity amplitudes of both stars. 

For each light curve, we add an overall scaling parameter, and for the CHIMERA data, a third-order polynomial to model any long-timescale trends in the data. In addition, each light curve has a parameter that adds additional uncertainty in the case the uncertainties are underestimated. We also added such a parameter to the entire spectral energy distribution.

Finally, binary orbits can be eccentric, which introduces two additional parameters: eccentricity ($e$) and the longitude of periastron ($\omega$). This will result in a small time offset ($\Delta t_{e}$) of the secondary eclipse compared to phase 0.5 \citep{winn2010}:
\begin{equation}
    \Delta t_{e} \approx \dfrac{P}{\pi} e \cos \omega 
\end{equation}

Because white dwarfs are massive, compact objects, there is also a small time delay of the secondary eclipse as a result of relativistic light travel time delay, the R{\o}mer delay \citep{kaplan2010,kaplan2014}:
\begin{equation}
\label{eq:dtLT}
    \Delta t_{LT} = \dfrac{P K_1}{\pi c} (1-q^{-1}) 
\end{equation}
In our light curve model, the R{\o}mer delay is not included in the calculation of the light curve. When modelling the light curve, we assume that the time delay to the secondary eclipse is solely due to the eccentricity. In the final calculation of the parameters, we correct for the R{\o}mer delay and recalculate the eccentricity. 

Finally, we estimate how much the orbital period is expected to change due to gravitational waves angular momentum losses using:
\begin{equation}
    \dot{P}_\mathrm{GW} = \dfrac{96}{5} \left( 
\dfrac{G \mathcal{M} }{P} \right )^{5/3} \dfrac{2\pi}{c^5}
\end{equation}
where $\mathcal{M} = \dfrac{(M_1 M_2)^{3/5}}{(M_1 + M_2)^{1/5}}$, the chirp mass.

\subsubsection{Fitting process}
To find the best model and parameter ranges, we model light curves and SED at the same time using \textit{emcee}. We ran each chain at least \qty{10000}{} times and inspected the results. If the parameters did not reach a steady state, we continued the chains until the parameter distributions stabilized. 

The challenge here is that there are a large number of free parameters and large and different datasets that constrain them. In addition, there are many parameter correlations and some degeneracies. For example, in the case of grazing eclipses, there is a degeneracy between the radii and inclination.

The biggest uncertainty is the scale of the systems, the masses or semi-major axis. The light curve shape is only determined by the scaled radii: $r_\mathrm{1,2}=R_\mathrm{1,2}/a$. For four systems, we have atleast one radial velocity measurement to constrain the mass of the stars, and with that, the semi-major axis. However, for the three systems we do not have any constrain on the mass or size of the system.

To make sure that the solution are physically possible, we use a minimum and maximum radius for both stars. For the minimum size, we use the zero-temperature white dwarf mass-radius relation \citep{verbunt1988}. For the maximum size of the white dwarfs, we use the radius of a white dwarf with a temperature of \qty{40000}{K}\footnote{\url{https://evolgroup.fcaglp.unlp.edu.ar/public/oldpage/TRACKS/newtables.html}} \citep{althaus2013,camisassa2016,camisassa2019}. For high mass white dwarfs ($\gtrsim$\qty{0.6}{\Msun}), the different between these two limits is small (just a few percent). For lower mass white dwarfs, the difference between a cold and hot white dwarf can be a factor of 2 \citep[e.g.][]{parsons2018}. For consistency, we use these priors when fitting all systems, including systems for which we have radial velocity constraints. 

Figures \ref{fig:overview_ZTFJ0221}--\ref{fig:overview_ZTFJ2249} show the model fits to the $g$, $r$, and $i$ light curves (both ZTF and CHIMERA) and the spectral energy distribution. The figures show that the models fit the light curves well. The model also fit the SED well, although in some cases there is a small discrepancy between the prediction and measurements in the UV bands (FUV, NUV).


\section{Results}\label{sec:results}
We measured the binary parameters of the seven white dwarf binaries we found in the ZTF data, see Table~\ref{tab:WDparameters}. The orbital periods of the white dwarfs is between 45 minutes and 3 hours. The primary white dwarf (the most luminous) are the hottest white dwarf in the binary, with temperatures between \qtyrange{9000}{19000}{K}. The secondary white dwarfs are colder with temperatures of \qtyrange{4300}{12000}{K}. The masses of most white dwarfs range from \qtyrange{0.23}{0.45}{\Msun} and the mass ratio is typically slightly less than 1. There is one exception; the secondary white dwarf of ZTF~J0221+1710 has a higher mass of \qty{0.61}{\Msun} and has a mass ratio of 2.1. During the fitting procedure, we limited the radii of both white dwarfs using the mass-radius relation. Therefore, the reported radii fall within these limits. For systems for which we measured one or two radial velocity amplitudes, we have a tighter constraint on the mass and radii. For these systems, most masses and radii correspond to the prediction from the theoretical mass-radius-temperature models. However, there are some exceptions. For ZTF~J0238+0933 the secondary white dwarf is slightly larger than the model would predict, and for J1356+5705 the secondary white dwarf is predicted to be much smaller and lower mass. We also note that for ZTF~J0221+1710 was already characterised by \citet{kosakowski2023} and we find the same binary parameters. 

Six out of the seven systems have detected primary and secondary eclipses. For these systems, we measured the time offset of the secondary eclipse relative to orbital phase 0.5 ($\Delta t_2$). We can typically constrain this value to $\approx$\qty{1}{s}, but it depends on the depth of the secondary eclipse and the quality (and amount) of high-cadence photometry data. For the two longest period systems, ZTF~J0238+0933 and ZTF~J1758+7642, we measure a significant delay of the secondary eclipse of $\approx$10--12 seconds. We calculate the corresponding value for $e \cos \omega$ for both these stars which is about $10^{-3}$.

\begin{landscape}
\begin{table}
\small
\renewcommand{\arraystretch}{1.5}
\caption{Binary parameters of the seven eclipsing white dwarfs. $^p$ indicates a prior was used on that parameter.} \label{tab:WDparameters}
\begin{tabular}{l|lllllll}
Name & J0221+1710 & J0238+0933 & J0720+6439 & J1110+7445 & J1356+5705 & J1758+7642 & J2249+0117 \\
\hline\hline
$P$  ($\mathrm{BJD_{TDB}}$)   & $ 0.0612853495(8) $               & $ 0.136828990(4) $             & $ 0.0314116865(3) $               & $ 0.1205767140(10) $              & $ 0.0638397116(3) $               & $ 0.1313333951(10) $              & $ 0.0956575227(20) $              \\
$t_0$ ($\mathrm{BJD_{TDB}}$)  & $ 2458759.947545(9) $             & $ 2458782.86201(2) $           & $ 2458796.922867(5) $             & $ 2458435.948557(9) $             & $ 2458593.757803(4) $             & $ 2459762.698510(5) $             & $ 2459120.86164(2) $              \\
$M_1$ (\qty{}{\Msun})         & $ 0.262^{+0.017}_{-0.016} $    & $ 0.35^{+0.06}_{-0.06} $    & $ 0.31^{+0.05}_{-0.05} $       & $ 0.33^{+0.08}_{-0.09} $       & $ 0.45^{+0.03}_{-0.03} $       & $ 0.29^{+0.07}_{-0.08} $       & $ 0.24^{+0.08}_{-0.09} $       \\
$M_2$ (\qty{}{\Msun})         & $ 0.6150^{+0.0166}_{-0.0155} $    & $ 0.299^{+0.041}_{-0.048} $    & $ 0.274^{+0.024}_{-0.026} $       & $ 0.390^{+0.047}_{-0.082} $       & $ 0.267^{+0.018}_{-0.014} $    & $ 0.21^{+0.09}_{-0.07} $       & $ 0.37^{+0.06}_{-0.07} $       \\
$R_1$ (\qty{}{\Rsun})         & $ 0.0278^{+0.0005}_{-0.0004} $ & $ 0.017^{+0.003}_{-0.002} $ & $ 0.0246^{+0.0010}_{-0.0015} $ & $ 0.0164^{+0.0009}_{-0.0022} $    & $ 0.0149^{+0.0008}_{-0.0006} $ & $ 0.0189^{+0.0012}_{-0.0012} $ & $ 0.0216^{+0.0011}_{-0.0014} $ \\
$R_2$ (\qty{}{\Rsun})         & $ 0.01205^{+0.00025}_{-0.00018} $ & $ 0.022^{+0.003}_{-0.004} $ & $ 0.0186^{+0.0015}_{-0.0010} $ & $ 0.0165^{+0.0017}_{-0.0008} $ & $ 0.0181^{+0.0004}_{-0.0004} $ & $ 0.0211^{+0.0013}_{-0.0014} $ & $ 0.0165^{+0.0013}_{-0.0010} $ \\
$T_1$ (K)                     & $ 13090^{+120}_{-200} $           & $ 18440^{+1300}_{-1060} $      & $ 18200^{+630}_{-650} $           & $ 13820^{+630}_{-540} $           & $ 10130^{+280}_{-280} $           & $ 15190^{+730}_{-590} $           & $ 19010^{+560}_{-500} $           \\
$T_2$ (K)                     & $ 4200^{+570}_{-600} $            & $ 11800^{+680}_{-700} $        & $ 7620^{+270}_{-280} $            & $ 10240^{+290}_{-250} $           & $ 8440^{+180}_{-200} $            & $ 12530^{+540}_{-470} $           & $ 8880^{+250}_{-240} $            \\
$\log g_1$ [cgs]              & $ 6.95^{+0.02}_{-0.03} $       & $ 7.33^{+0.16}_{-0.14} $    & $ 7.13^{+0.04}_{-0.04} $       & $ 7.41^{+0.17}_{-0.10} $       & $ 7.70^{+0.04}_{-0.04} $       & $ 7.30^{+0.07}_{-0.10} $       & $ 7.12^{+0.09}_{-0.11} $       \\
$\log g_2$ [cgs]              & $ 8.04^{+0.02}_{-0.02} $       & $ 7.06^{+0.13}_{-0.08} $    & $ 7.26^{+0.05}_{-0.06} $       & $ 7.49^{+0.07}_{-0.14} $       & $ 7.32^{+0.03}_{-0.03} $       & $ 7.08^{+0.10}_{-0.10} $       & $ 7.49^{+0.08}_{-0.10} $       \\
$r_1$                         & $ 0.0439^{+0.0008}_{-0.0007} $ & $ 0.016^{+0.003}_{-0.003} $ & $ 0.0689^{+0.0017}_{-0.0024} $    & $ 0.0167^{+0.0006}_{-0.0024} $    & $ 0.02442^{+0.00104}_{-0.00085} $ & $ 0.02051^{+0.00119}_{-0.00067} $ & $ 0.02763^{+0.00076}_{-0.00111} $ \\
$r_2$                         & $ 0.0189^{+0.0004}_{-0.0003} $ & $ 0.022^{+0.002}_{-0.003} $ & $ 0.051^{+0.004}_{-0.003} $    & $ 0.0170^{+0.0019}_{-0.0006} $ & $ 0.0295^{+0.0005}_{-0.0008} $ & $ 0.0232^{+0.0005}_{-0.0009} $ & $ 0.0205^{+0.0017}_{-0.0012} $ \\
$K_\mathrm{1}$ (\qty{}{km/s}) & $ 359^{+4}_{-4} $           & $ 157^{+13}_{-14} $      & $ 256.6^{+11}_{-11} $         & $ 188^{+24}_{-33} $               & $ 172.4^{+7}_{-8} $           & $ 125^{+38}_{-31} $               & $ 213^{+30}_{-32} $               \\
$K_\mathrm{2}$ (\qty{}{km/s}) & $ 153^{+8}_{-6} $           & $ 187.7^{+17}_{-17} $      & $ 296^{+20}_{-25} $               & $ 160^{+32}_{-34} $               & $ 296.5^{+10}_{-10} $         & $ 172^{+30}_{-35} $               & $ 142^{+34}_{-42} $               \\
age1 (Gyr)                    & $ 0.091^{+0.029}_{-0.020} $       & $ 0.039^{+0.030}_{-0.025} $    & $ 0.021^{+0.027}_{-0.054} $       & $ 0.137^{+0.061}_{-0.066} $       & $ 0.426^{+0.108}_{-0.066} $       & $ 0.067^{+0.049}_{-0.077} $       & $ -0.003^{+0.050}_{-0.034} $      \\
age2 (Gyr)                    & $ 4.4^{+3.4}_{-3.0} $             & $ 0.205^{+0.108}_{-0.106} $    & $ 1.04^{+0.23}_{-0.19} $          & $ 0.409^{+0.197}_{-0.072} $       & $ 0.747^{+0.100}_{-0.086} $       & $ 0.073^{+0.178}_{-0.062} $       & $ 0.62^{+0.22}_{-0.24} $          \\
$a$  (\qty{}{\Rsun})          & $ 0.627^{+0.007}_{-0.006} $    & $ 0.97^{+0.04}_{-0.04} $    & $ 0.351^{+0.011}_{-0.014} $    & $ 0.954^{+0.028}_{-0.025} $       & $ 0.605^{+0.012}_{-0.009} $    & $ 0.90^{+0.04}_{-0.05} $       & $ 0.77^{+0.03}_{-0.03} $       \\
$i$ (deg)                     & $ 88.98^{+0.16}_{-0.15} $      & $ 88.90^{+0.07}_{-0.04} $   & $ 87.6^{+0.4}_{-0.4} $         & $ 89.72^{+0.14}_{-0.10} $      & $ 89.51^{+0.04}_{-0.03} $      & $ 89.67^{+0.04}_{-0.02} $      & $ 89.03^{+0.17}_{-0.18} $      \\
$q$                           & $ 2.15^{+0.11}_{-0.12} $       & $ 0.72^{+0.14}_{-0.11} $    & $ 0.76^{+0.09}_{-0.07} $       & $ 0.8^{+0.4}_{-0.3} $          & $ 0.55^{+0.04}_{-0.04} $       & $ 0.5^{+0.3}_{-0.2} $          & $ 1.0^{+0.6}_{-0.3} $          \\
$e \times 10^{-3}$                           & 0                                 & $ 1.75^{+0.15}_{-0.15} $    & <6.2              & <0.7             & <0.5             & $ 1.55^{+0.08}_{-0.08} $       & <1.5              \\
distance (pc)                 & $ 321^{+5}_{-7} $           & $ 641^{+28}_{-32} $            & $ 752^{+28}_{-36} $               & $ 427^{+16}_{-14} $         & $ 337^{+8}_{-8} $           & $ 650^{+36}_{-36} $               & $ 556^{+23}_{-26} $               \\
Egr (mag)                     & $ 0.117^{+0.006}_{-0.007} $    & $ 0.056^{+0.011}_{-0.010} $ & $ 0.053^{+0.012}_{-0.012} $    & $ 0.057^{+0.013}_{-0.013} $    & $ 0.066^{+0.017}_{-0.016} $    & $ 0.057^{+0.013}_{-0.014} $    & $ 0.071^{+0.012}_{-0.012} $    \\
$\Delta t_2$ (s)              & 0                                 & $ 12.7^{+1.2}_{-1.1} $      & $ -0.5^{+3.3}_{-1.4} $            & $ 0.3^{+1.4}_{-1.5} $          & $ -1.5^{+0.7}_{-0.6} $         & $ 10.5^{+0.8}_{-0.7} $         & $ -0.6^{+2.8}_{-0.9} $            \\
$\Delta t_\mathrm{LT}$ (s)    & $ 1.08^{+0.05}_{-0.06} $       & $ -0.8^{+0.4}_{-0.4} $      & $ -0.23^{+0.09}_{-0.08} $      & $ -0.4^{+0.6}_{-0.7} $         & $ -0.84^{+0.09}_{-0.09} $      & $ -1.3^{+0.8}_{-0.7} $         & $ 0.0^{+0.6}_{-0.5} $         \\
$\tau_\mathrm{merge}$ (Myr)   & $ 145^{+10}_{-9} $          & $ 1280^{+380}_{-240} $         & $ 32^{+8}_{-5} $            & $ 755^{+157}_{-130} $             & $ 193^{+14}_{-18} $         & $ 1400^{+530}_{-440} $            & $ 476^{+178}_{-113} $  \\
$\dot{P}_\mathrm{GW}$ (s/s) $\times 10^{-15}$ &	$ 18.1^{+01.2}_{-01.2} $ &	$3.5^{+0.8}_{-0.9} $ &	$ 33.7^{+6.1}_{-7.4} $&	$ 05.7^{+1.1}_{-01.0} $ &	$ 13.8^{+01.4}_{-0.9} $&$ 2.7^{+1.1}_{-0.8} $&$6.1^{+1.9}_{-1.8} $

\end{tabular}
\end{table}
\end{landscape}

\section{Discussion}\label{sec:discussion}

\subsection{Binary properties and evolution}
Given the masses of the binaries, we conclude that most are double helium-core white dwarfs. These systems evolve from two relatively low-mass main-sequence binaries ($\lesssim$ \qty{2.3}{\Msun}) in orbits of a few tens of days. During the first and second mass transfer phases, the mass transfer starts before helium core ignition. The first stage of mass-transfer is unstable, but because of the equal mass ratio, the envelope is ejected without the orbital period significantly decreasing. The second mass transfer phase is also unstable, and because of unequal masses, results in a common envelope that shrinks to orbit to a few hours as we see it today \citep{nelemans2001a}. The exception is ZTF~J0221+1710, is has a small and more massive secondary white dwarf and is most likely a He-CO double white dwarf. It started out as a more unequal mass binary and experienced a common envelope during the first phase of mass transfer. As the lower mass star ascends the first giant branch, it evolves into a helium white dwarf.

These white dwarf binaries have similar masses and radii as compared to the shorter period eclipsing white dwarfs found using ZTF \citep{burdge2020}. The majority are double Helium white dwarfs systems with relatively low masses, with a few He-CO binaries with more unequal mass-ratios. We do note that the longer period systems presented in this paper tend to have lower temperatures and are intrinsically fainter (see Fig.~\ref{fig:HRdiagram}). The new systems have distances of \qtyrange{380}{970}{parsec}, and are closer than almost all the short period eclipsing dWDs found by \citet{burdge2020} (\qtyrange{640}{6000}{parsec}). This implies that, even though the number of ZTF discovered systems with periods shorter and longer than one hour is roughly the same, the intrinsic occurrence rate of long period dWDs is far higher, and the long period systems are more representative of typical double white dwarfs. 
The fact the majority of binaries have atleast one low mass component is likely a selection effect (brighter, higher eclipse propability), and is unlikely to reflect the intrinsic population of double white dwarf binaries (see also Section \ref{sec:future}).

\subsection{Eccentricity}\label{sec:eccentricity}
\begin{figure*}
    \centering
    \includegraphics{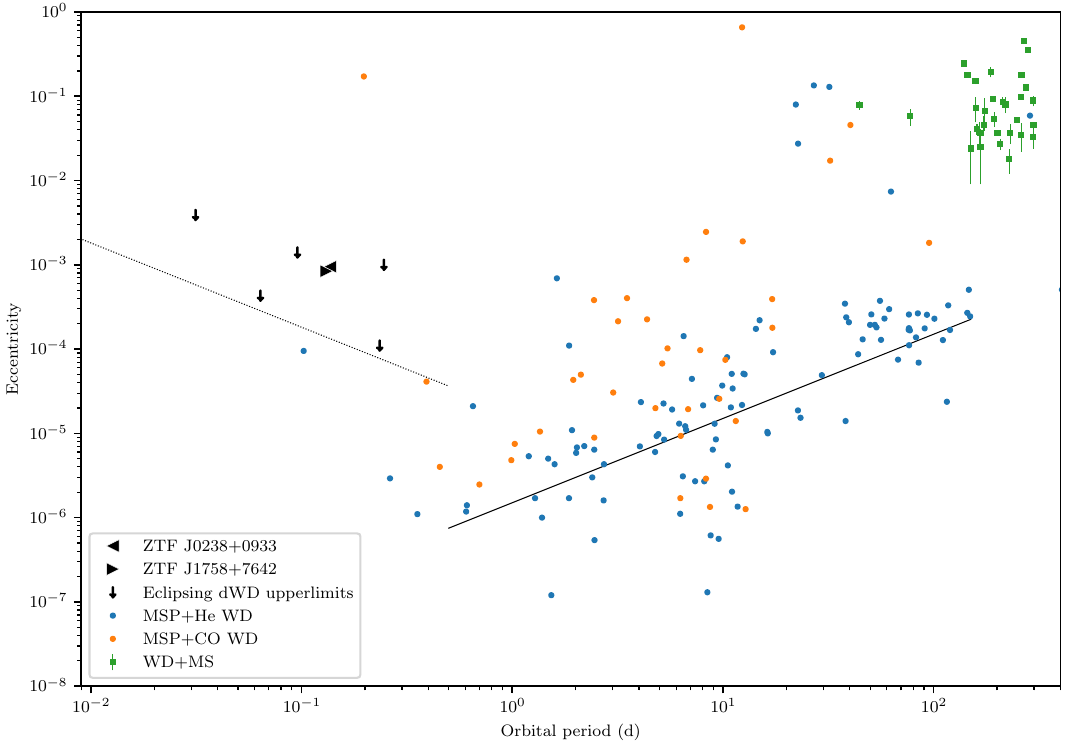}
    \caption{Orbital period versus the eccentricity of binaries that contain white dwarf stars. Note that for the eclipsing white dwarfs, we can only constrain $e \cos \omega$ (and not $e$), and therefore we should consider these points as lower limits. The arrows indicate upperlimits for $e\cos\omega$ from the eclipsing white dwarfs in this paper and from \citet{hermes2014} and \citet{kaplan2014}. The white dwarfs with millisecond pulsars are taken from the ATNF Pulsar catalogue \citep{manchester2005}, with pulsars from globular clusters removed. The long period white dwarf main sequence binaries are from \citet{yamaguchi2024}.
    The dashed line shows the detection limit of eccentricity in eclipsing double white dwarfs assuming a 1 second precision on the eclipse arrival time. The solid line show the theoretical prediction for MSP+WD binaries that are the result of stable mass-transfer \citep{phinney1992}.}
    \label{fig:e}
\end{figure*}

As can be seen in Table~\ref{tab:WDparameters}, the two longest-period systems (ZTF~J0238+0933 and ZTF~J1758+7642) show a significant delay in the secondary eclipse arrival time. Note that we detected this in multiple nights of CHIMERA data and in multiple filters. An offset of the secondary eclipse can the result of the R{\o}mer delay, but for these two binaries this delay is small (because $q\approx1$). For ZTF~J0238+0933 and ZTF~J1758+7642, the secondary eclipse arrival time is $\approx$\qtyrange{10}{12}{s} late, and we conclude that this delay is due to a small eccentricity of the binary. Note that we cannot constrain the longitude of periastron ($\omega$), and therefore we can only determine a lower limit to the eccentricity (Eq. \ref{eq:dtLT}). For both systems, this is approximately $e \cos \omega \approx 2 \times 10^{-3}$.

This is the first time a non-zero eccentricity is measured for a short period double white dwarf binary and short period post common envelope binary in general. As shown in Fig.~\ref{fig:e}, there are longer period white dwarf binaries that are eccentric, and there are also many millisecond pulsars with white dwarf companions that are eccentric. In these cases, the binaries likely formed through stable mass transfer, and the eccentricity is a result of that process \citep{phinney1992}. There are also longer period (100--1000 days) white dwarf -- main sequence binaries found using \textit{Gaia} with eccentricities of $\approx$0.1 \citep{yamaguchi2024} or even larger \citep{shahaf2024}.

It was often assumed that the common envelope would rapidly circularise binaries, but there is some recent theoretical work on common envelopes that suggests otherwise. \citet{sand2020,glanz2021,bronner2024}. They show that, using hydrodynamical simulations, that even if the pre- common envelope binary is circular, it can still result in a post common envelope binary with an eccentricity of $e\approx 0.03$. This is within an order of magnitude of what we find.

If we assume that double white dwarf binaries exit the common envelope slightly eccentric, they can still circularize. In the case of our two binaries, tidal interaction can be neglected because the relative size of the white dwarfs ($r_\mathrm{1,2}$) are small \citep{campbell1984, hurley2002}. Gravitational wave losses are more important. Using Equation 5.7 from \citet{peters1964}, we calculate $\dot e$ and estimate the circularization timescale due to gravitational waves. This is $\approx$\qty{2.5}{Gyr}, an order of magnitude larger than the time since the common envelope. We also calculate this timescale for the shorter period systems that are closer together, and for those, the circularization timescale become similar to the time since common envelope. This, and the lower sensitivity at shorter periods, would explain why we detected a small eccentricity for the longer period systems and not for any of the shorter period double white dwarf binaries. 

If these binaries are indeed eccentric, we can expect the longitude of periastron ($\omega$) to slowly change due to the relativistic apsidal precession. 
We can calculate the relativistic apsidal precession using the equation from \citet{gimenez1985},
\begin{equation}
\Dot{\omega} = 5.447\times 10^{-4} \dfrac{(M_1+M_2)^{2/3}}{(1-e^2)P^{2/3}} \mathrm{deg/cycle}.
\end{equation}
where the masses are in solar units and orbital period in days.
If we assume $e=10^{-3}$, the relativistic apsidal precession is $\Dot{\omega}\approx$ \qty{0.7}{deg/year} for both systems, which corresponds to an apsidal precession period of 460 years.

We note that the eccentricity could also be the result of triple interactions. If there is a tertiary object in the system, Kozai-Lidov oscillations can result in small eccentricities in the inner binary. However, there is no sign of any third object in the spectra of spectral energy distribution, so if there is a third object, it would have to be a brown dwarf or large planet.

\subsection{Future evolution and gravitational waves}
All these white dwarf binaries emit gravitational waves and slowly spiral inward. We can calculate the time until a merger using:
\begin{equation}\label{eq:GWtimescale}
    \tau_\mathrm{merge} = \dfrac{5}{256} \dfrac{a^4 c^5}{G^3 M_1 M_2 (M_1+M_2)}
\end{equation}
Using this equation, we calculate that the time to merger takes \qty{30}{Myr} to \qty{2.2}{Gyr}. The result of the merger depends on the mass and mass-ratio \citep{shen2015}. The double helium white dwarfs we presented in this work will likely merge into a subdwarf stars and later evolve into a single white dwarf \citep{han2002,marsh2004}. ZTF~J0221+1710 is a CO-He system and will merge to became an R Coronae Borealis (R~CrB) star \citep{webbink1984,iben1984,saio2002}  and eventually evolve into a CO white dwarf.

The most nearby and shortest period double white dwarf binaries will be directly detectable by the LISA gravitational wave detector. We used the binary parameters and estimate the expected SNR for the LISA satellite \citep{wagg2022}. The source with the highest SNR is ZTF~J1110+7445 (the shortest period system), which is predicted to have a SNR in LISA of 4.4 after 10 years. Because the sky-location and frequency is known, it might be detectable as an individual sources after LISA collected data for 10 years. The other systems are predicted to have an SNR of 1-2, and therefore will contribute to the unresolved `foreground' signal of LISA.

\section{Future work}\label{sec:future}
High-speed cameras on large telescopes have the potential to measure eclipse times to a precision of $\approx$\qty{0.5}{s} (e.g. HiPERCAM, \citealt{dhillon2021}). With this precision, we can expect a measurable change in the arrival time of the eclipse in the next decade for three systems. For ZTF~J0238+0933 and ZTF~J1758+7642, we measured a delay in the arrival time of the secondary eclipse compared to the primary eclipse ($\Delta t_2$) that we attribute to a small eccentricity in the orbit of the two stars. Due to the relativistic apsidal precision, the orientation of the orbit will change with $\approx 0.7$ deg/yr (see Sect. \ref{sec:eccentricity}). If we assume that the current orientation is $\omega=45\deg$, this would result in a measurable change of $\Delta t_2\approx 1.4$s in 10 years. By measuring the amplitude (and period) of the secondary eclipse arrival time changes, we can constrain the eccentricity and further constrain the masses of both stars. 

In addition, we can also expect a change in the eclipse arrival time because of angular momentum loss due to gravitational waves. The change in eclipse arrival time is:
\begin{equation}
    \Delta t_\mathrm{GW} = \dfrac{\dot{P_\mathrm{GW}}}{2 P} t^2
\end{equation}
However, because the orbital periods are relativly long for most systems, we can only expect to measure this effect for the shortest period system; ZTF~J0720+6439. For this binary, we predict a change in the eclipse arrival time of \qty{0.6}{s} in 10 years, and \qty{1.0}{s} in 13 years. Measurements of the exact change can be used to determine the precise masses of the two stars \citep{burdge2019a}.

Finally, weworth monitoring the eclipse arrival time of all these systems to probe the effect of orbital decay due to tides \citep[e.g.][]{benacquista2011, piro2011, fuller2012, piro2019}.


Between the search performed by \citet{burdge2020} and this paper, we systematically searched for and found all eclipsing double white dwarfs in the ZTF data brighter than 19 mag. This brings the total number of known eclipsing dWDs to 23, of which 20 are in the ZTF footprint\footnote{Two are in the South, and one is in a ZTF chipgap}. Because the selection biases of eclipses are easy to model, this sample of 20 double white dwarfs can be used to infer the intrinsic population properties of double white dwarfs (e.g. \citealt{maoz2024}, Van Zeist et al. 2025, in prep).

\section{Summary and conclusion}\label{sec:conclusion}

We summarize the paper as follows:
\begin{itemize}
    \item We discovered or recovered seven eclipsing double white dwarf binaries in a systematic search of ZTF data for eclipsing white dwarfs. The binaries have orbital period of 45 minutes to 3 hours, are between are typically $G$ 18.5--19 mag with one brighter object a 17.6, and have distances between 270 to \qty{900}{parsec}.
    This brings the total number of known eclipsing double white dwarfs to 26, significantly increases the number of nearby eclipsing double white dwarfs, and doubles the number of systems with orbital periods longer than 1 hour.
    \item We obtained high-speed photometry and ID-spectra for all seven eclipsing double white dwarfs of both the primary and secondary eclipses. We also obtained phase-resolved spectra for four systems. Using this data, we measured the binary parameters by modelling the archival multi-wavelength data, all light curves, and the radial velocity measurements. 
    \item Most of the white dwarfs are roughly equal mass binaries with low-mass, helium cores. One system has a higher mass secondary and his a He-CO double white dwarf binary. All systems have a DA-spectral type. The masses and mass-ratios of these eclipsing double white dwarfs are similar to previously known eclipsing binary white dwarfs.
    \item Two systems show a significant delay in the secondary eclipse time. We conclude that this is due to a small eccentricity in the orbit of about $2\times10^{-3}$, the first time an eccentricity has been measured for a compact double white dwarf system. We suggest that these systems emerged from the common envelope with this small eccentricity, and gravitational waves did not have the time to fully circularize the orbit. We predict that, due to relativistic apsidal precession,  there will be a detectable change in the delay of the secondary eclipse within a decade.
    \item All seven systems are losing angular moment due to gravitational wave radiation. They will get into contact in 36--1200 Myr. Based on their masses and mass ratios, we predict that these systems will merge and finally evolve into a single CO white dwarf. For the shortest period system, we expect a measurable eclipse arrival time within a decade due to gravitational wave angular momentum losses. In addition, LISA might be able to detect its gravitational waves directly with an expected SNR of $\approx$4.
\end{itemize}

\begin{acknowledgements}

We thank Wouter van Zeist, Gijs Nelemans, and James Munday for useful conversations.

This publication is part of the project "The life and death of white dwarf binary stars" (VI.Veni.212.201) of the research programme NWO Talent Programme Veni Science domain 2021 which is financed by the Dutch Research Council (NWO).

This research was supported by Deutsche Forschungsgemeinschaft  (DFG, German Research Foundation) under Germany’s Excellence Strategy - EXC 2121 "Quantum Universe" – 390833306. Co- funded by the European Union (ERC, CompactBINARIES, 101078773) Views and opinions expressed are however those of the author(s) only and do not necessarily reflect those of the European Union or the Euro- pean Research Council. Neither the European Union nor the granting authority can be held responsible for them

Based on observations obtained with the Samuel Oschin Telescope 48-inch and the 60-inch Telescope at the Palomar Observatory as part of the Zwicky Transient Facility project. ZTF is supported by the National Science Foundation under Grants No. AST-1440341 and AST-2034437 and a collaboration including current partners Caltech, IPAC, the Oskar Klein Center at Stockholm University, the University of Maryland, University of California, Berkeley, the University of Wisconsin at Milwaukee, University of Warwick, Ruhr University, Cornell University, Northwestern University and Drexel University. Operations are conducted by COO, IPAC, and UW.

This work used the Palomar Observatory 5.0m Hale telescope, which is owned and operated by the Caltech Optical Observatories.

This research has made use of the SIMBAD database, operated at CDS, Strasbourg, France. 

\end{acknowledgements}

%
%

\bibliographystyle{aa}
\bibliography{references} 

\begin{appendix}

\section{Extra Tables and Figures}
\begin{figure}
    \centering
    \includegraphics[]{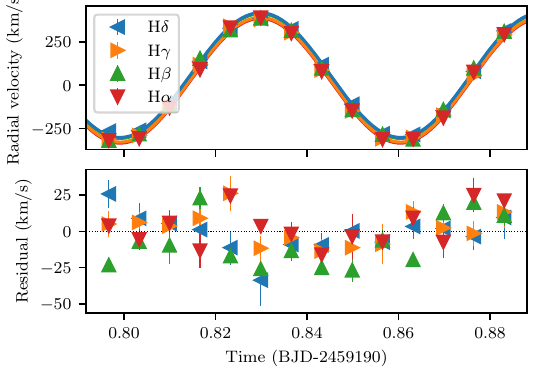}
    \caption{Radial velocity measurements of ZTF~J0221+1710 measured from the ESI data. The different colours indicate velocity measured from three different lines. }
    \label{fig:RVcurve_ZTFJ0221_ESI}
\end{figure}

\begin{figure}
    \centering
    \includegraphics[]{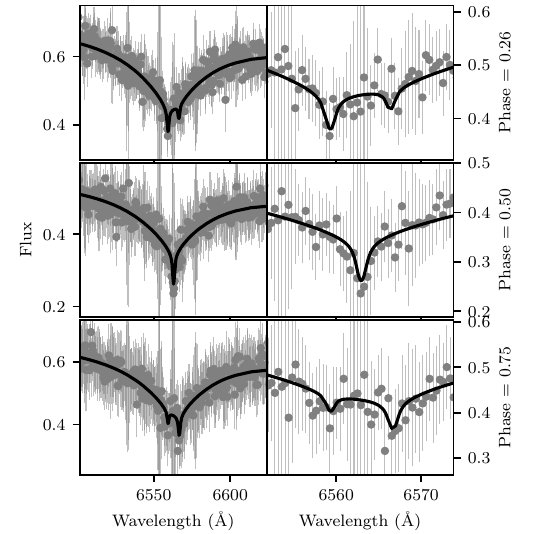}
    \caption{Spectra and fit of three spectra at different phases of ZTF~J0238+0933. It shows that the dominant line move as function of phase. Although we model a second component, the velocity amplitude ($K_2$) is not constrained well by the data.}
    \label{fig:spectrafit_ZTFJ0238}
\end{figure}

\begin{figure}
    \centering
    \includegraphics[]{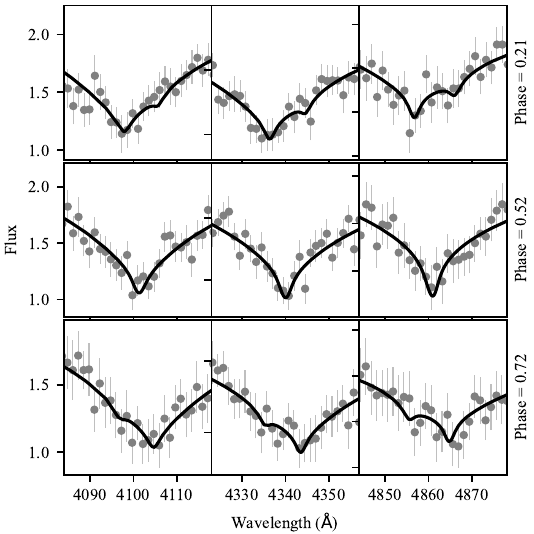}
    \caption{Spectra and fit of three spectra at different phases of ZTF~J0720+6439. It shows that the strongest components is clearly moving, but the secondary white dwarf is only barely detected.}
    \label{fig:spectrafit_ZTFJ0720}
\end{figure}

\begin{figure}
    \centering
    \includegraphics[]{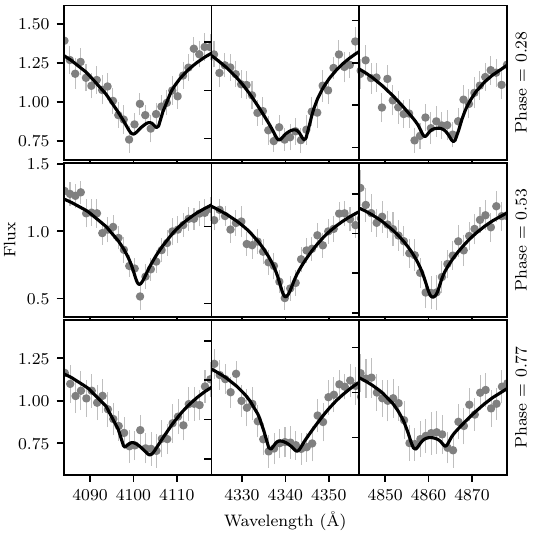}
    \caption{Spectra and fit of three spectra at different phases of ZTF~J1356+5705. It clearly shows that the absorption lines are composed of two components that are red-shifted and blue-shifted depending on the phase.}
    \label{fig:spectrafit_ZTFJ1356}
\end{figure}

\begin{figure*}
    \centering
    \includegraphics[]{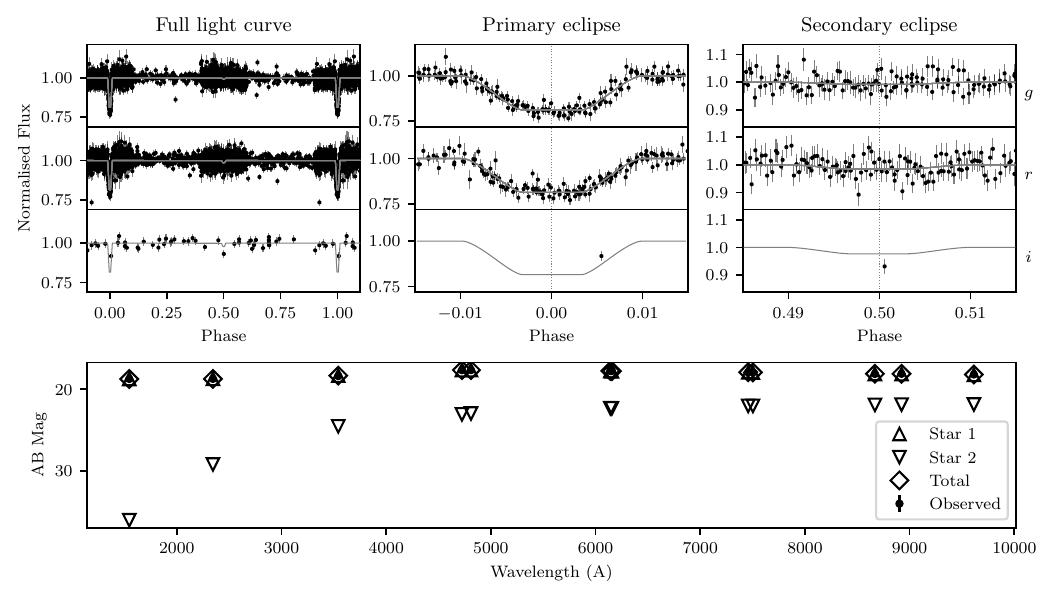}
    \caption{The joint fit to the light curves and spectral energy distribution of ZTF~J0221+1710.}
    \label{fig:overview_ZTFJ0221}
\end{figure*}

\begin{figure*}
    \centering
    \includegraphics[]{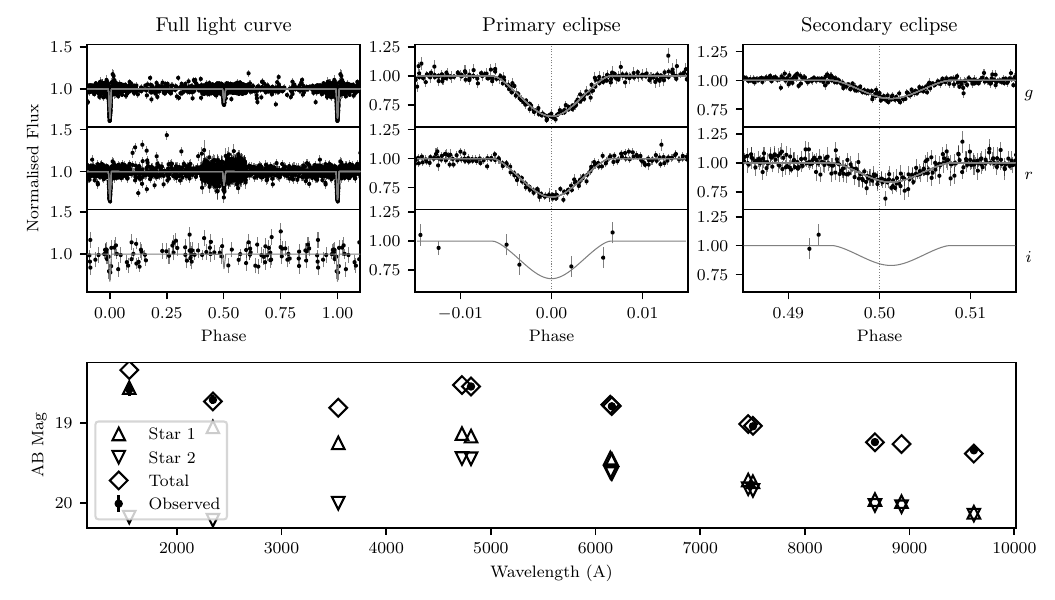}
    \caption{The joint fit to the light curves and spectral energy distribution of ZTF~J0238+0933.}
    \label{fig:overview_ZTFJ0238}
\end{figure*}

\begin{figure*}
    \centering
    \includegraphics[]{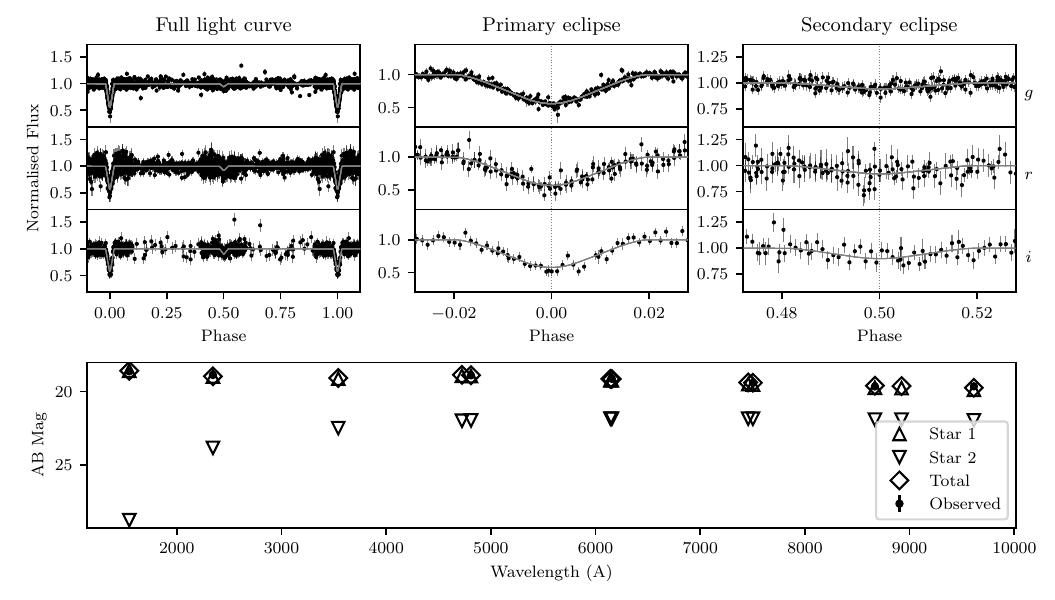}
    \caption{The joint fit to the light curves and spectral energy distribution of ZTF~J0720+6439.}
    \label{fig:overview_ZTFJ0720}
\end{figure*}

\begin{figure*}
    \centering
    \includegraphics[]{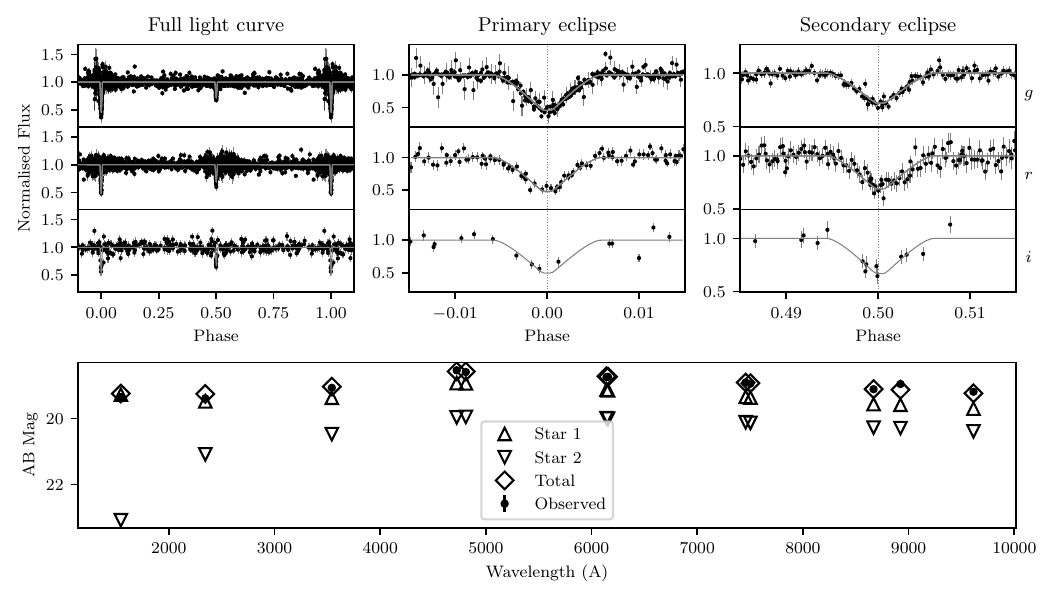}
    \caption{The joint fit to the light curves and spectral energy distribution of ZTF~J1110+7445.}    
    \label{fig:overview_ZTFJ1110}
\end{figure*}

\begin{figure*}
    \centering
    \includegraphics[]{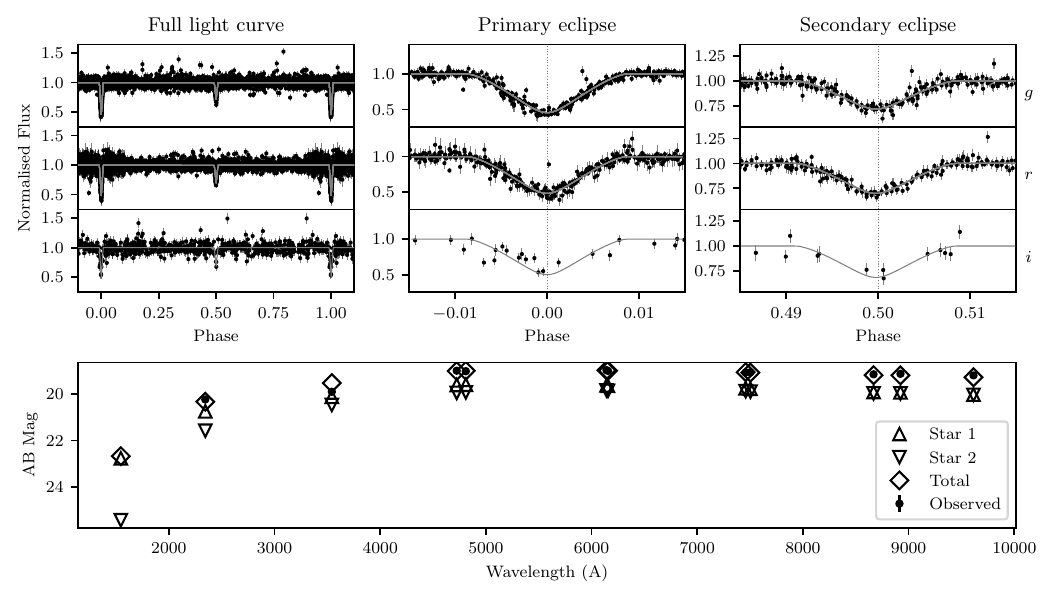}
    \caption{The joint fit to the light curves and spectral energy distribution of ZTF~J1356+5706.}
    \label{fig:overview_ZTFJ1356}
\end{figure*}

\begin{figure*}
    \centering
    \includegraphics[]{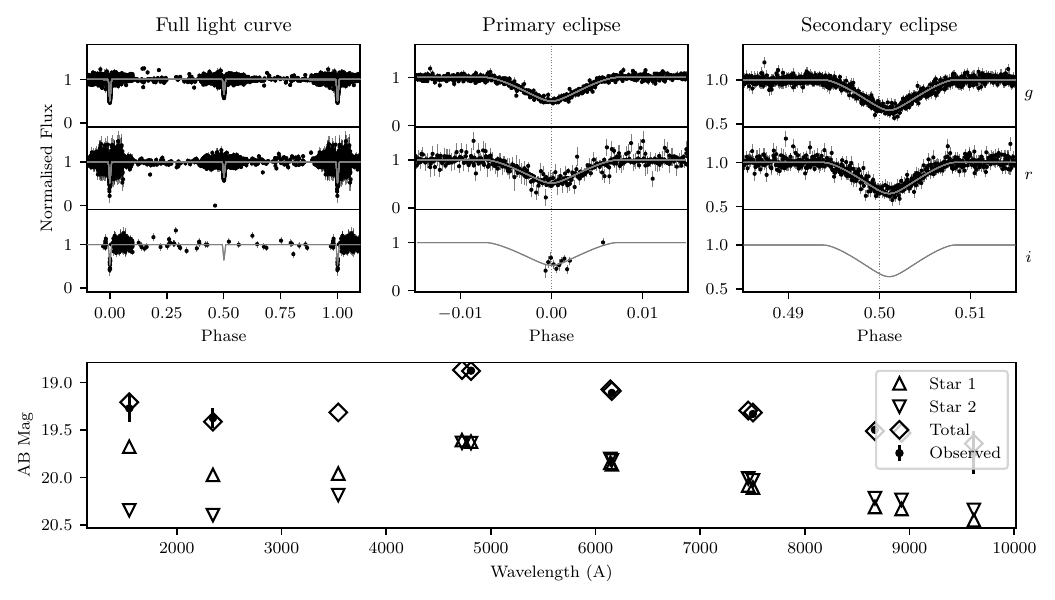}
    \caption{The joint fit to the light curves and spectral energy distribution of ZTF~J1758+7642.}
    \label{fig:overview_ZTFJ1758}
\end{figure*}

\begin{figure*}
    \centering
    \includegraphics[]{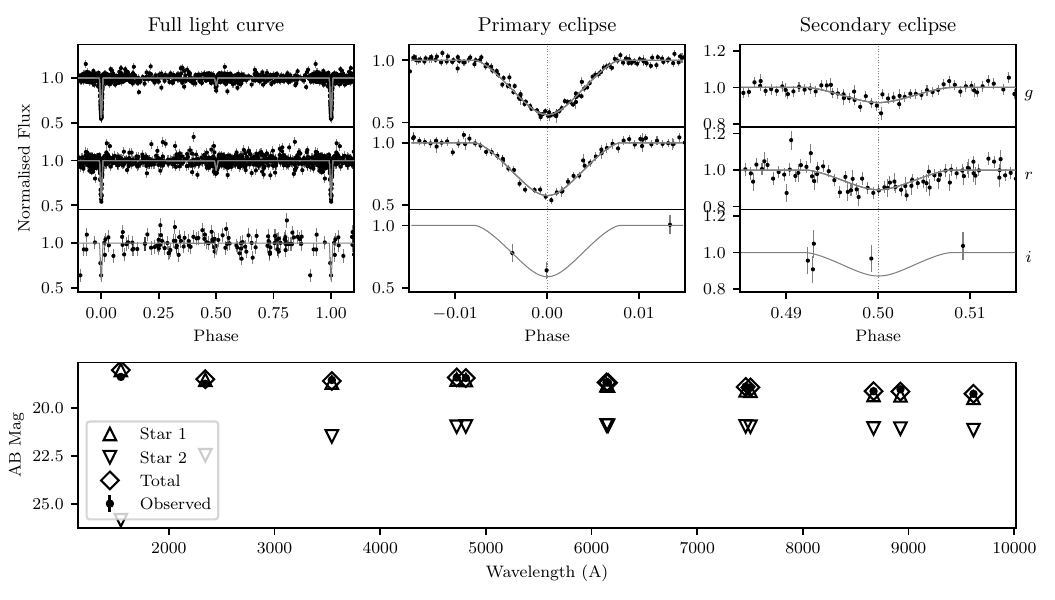}
    \caption{The joint fit to the light curves and spectral energy distribution of ZTF~J2249+0117.}
    \label{fig:overview_ZTFJ2249}
\end{figure*}

Figures \ref{fig:RVcurve_ZTFJ0221_ESI}, \ref{fig:spectrafit_ZTFJ0238}, \ref{fig:spectrafit_ZTFJ0720}, \ref{fig:spectrafit_ZTFJ1356} show fits to the phase resolved spectra. Tables \ref{tab:followup} and \ref{tab:CHIMERA} list followup observations. Figures \ref{fig:overview_ZTFJ0221}--\ref{fig:overview_ZTFJ2249} show the fit to the light curves and spectral energy distribution for each target.

\FloatBarrier

\begin{table*}[]
    \centering
    \begin{tabular}{l|lllll}
        Target & Telescope & Date & Exptime & Exposures & Setup \\
        \hline
        ZTF~J0221+1710 & ESI/Keck & 2020-09-16 & 520s & 14 &  Echellette \\ 
        ZTF~J0221+1710 & LRIS/Keck & 2020-12-16 & 300s & 14/16 & 600/4000 \& 600/7500 \\
        \rule{0pt}{2ex}ZTF~J0238+0933 & LRIS/Keck & 2022-02-02 & 600s & 4 & 600/4000 \& 400/8500 \\
        ZTF~J0238+0933 & LRIS/Keck & 2022-08-28 & 600s & 7/11 & 600/4000 \& 1200/7500  \\
        \rule{0pt}{2ex}ZTF~J0720+6439 & LRIS/Keck & 2022-03-07 & 240s & 11/10 & 600/4000 \& 600/7500  \\
        \rule{0pt}{2ex}ZTF~J1110+7445 & LRIS/Keck & 2022-03-07 & 300/250s & 4 & 600/4000 \& 600/7500  \\
        \rule{0pt}{2ex}ZTF~J1356+5706 & LRIS/Keck & 2020-09-16 & 300/250s & 23 & 600/4000 \& 600/7500  \\
        \rule{0pt}{2ex}ZTF~J1758+7642 & LRIS/Keck & 2022-08-28 & 900s & 5 & 600/4000 \& 1200/7500 \\       
        \rule{0pt}{2ex}ZTF~J2249+0117 & LRIS/Keck & 2024-11-08 & 500s & 1 & 600/4000 \& 400/8500  \\
    \end{tabular}
    \caption{An overview of all collected data. All LRIS data was obtained with a 1\arcsec slit. }
    \label{tab:followup}
\end{table*}

\begin{table*}
    \centering
    \begin{tabular}{ll|lll|lll}
\hline\hline
Target        & Date       & Filter & Nexp & Exptime & Filter & Nexp & Exptime   \\
\hline
ZTFJ0221+1710 & 2020-12-15 & g      & 345  & 3.0       & r & 348 & 3.0  \\
ZTFJ0221+1710 & 2020-12-15 & g      & 345  & 3.0       & r & 100 & 3.0  \\
ZTFJ0221+1710 & 2020-12-15 & g      & 350  & 3.0       & r & 345 & 3.0  \\
ZTFJ0221+1710 & 2020-12-15 & g      & 350  & 3.0       & r & 345 & 3.0  \\
ZTFJ0221+1710 & 2020-12-15 & g      & 99   & 3.0       & r & 93  & 3.0  \\
ZTFJ0221+1710 & 2022-08-24 & g      & 211  & 5.0       & r & 211 & 5.0  \\
\rule{0pt}{2ex}ZTFJ0238+0933 & 2022-01-04 & g      & 465  & 5.0       & r & 465 & 5.0  \\
ZTFJ0238+0933 & 2022-01-04 & g      & 400  & 5.0       & r & 400 & 5.0  \\
ZTFJ0238+0933 & 2022-01-05 & g      & 437  & 5.0       & r & 435 & 5.0  \\
ZTFJ0238+0933 & 2022-01-06 & g      & 427  & 5.0       & r & 426 & 5.0  \\
ZTFJ0238+0933 & 2022-01-06 & g      & 200  & 5.0       & r & 200 & 5.0  \\
ZTFJ0238+0933 & 2022-11-26 & g      & 150  & 5.6       & r & 200 & 5.6 \\
\rule{0pt}{2ex}ZTFJ0720+6439 & 2020-01-22 & g      & 87   & 3.0       & r & 89  & 3.0  \\
ZTFJ0720+6439 & 2020-01-22 & g      & 180  & 3.0       & r & 180 & 3.0  \\
ZTFJ0720+6439 & 2020-01-22 & g      & 180  & 3.0       & r & 180 & 3.0  \\
ZTFJ0720+6439 & 2020-01-22 & g      & 180  & 3.0       & r & 180 & 3.0  \\
ZTFJ0720+6439 & 2020-01-22 & g      & 35   & 3.0       & r & 90  & 3.0  \\
ZTFJ0720+6439 & 2020-01-23 & g      & 178  & 3.0       & r & 18  & 6.0   \\
ZTFJ0720+6439 & 2020-01-23 & g      & 180  & 3.0       & r & 90  & 6.0   \\
ZTFJ0720+6439 & 2022-02-28 & g      & 126  & 3.0       &   &     &     \\
ZTFJ0720+6439 & 2022-02-28 & g      & 178  & 3.0       & i & 178 & 3.0  \\
ZTFJ0720+6439 & 2022-02-28 & g      & 178  & 3.0       & i & 179 & 3.0  \\
\rule{0pt}{2ex}ZTFJ1110+7445 & 2020-01-22 & g      & 320  & 3.0       & r & 300 & 3.0  \\
ZTFJ1110+7445 & 2020-01-23 & g      & 525  & 3.0       & r & 519 & 3.0  \\
ZTFJ1110+7445 & 2020-08-19 & g      & 229  & 5.0       &   &     &     \\
ZTFJ1110+7445 & 2023-05-24 & g      & 300  & 5.0       & r & 300 & 5.0  \\
\rule{0pt}{2ex}ZTFJ1356+5706 & 2020-01-22 & g      & 366  & 3.0       & r & 314 & 3.0  \\
ZTFJ1356+5706 & 2020-01-23 & g      & 340  & 3.0       & r & 341 & 3.0  \\
ZTFJ1356+5706 & 2020-08-19 & g      & 129  & 5.0       &   &     &     \\
ZTFJ1356+5706 & 2020-08-19 & g      & 212  & 5.0       &   &     &     \\
ZTFJ1356+5706 & 2022-02-28 & g      & 109  & 10        & r & 109 & 10.0  \\
ZTFJ1356+5706 & 2022-02-28 & g      & 110  & 10        & r & 109 & 10.0  \\
ZTFJ1356+5706 & 2022-05-06 & g      & 363  & 3.0       & r & 363 & 3.0  \\
ZTFJ1356+5706 & 2022-05-06 & g      & 326  & 3.0       & r & 325 & 3.0  \\
ZTFJ1356+5706 & 2022-07-02 & g      & 84   & 3.0       & r & 84  & 3.0  \\
ZTFJ1356+5706 & 2022-07-02 & g      & 364  & 3.0       & r & 364 & 3.0  \\
ZTFJ1356+5706 & 2022-07-02 & g      & 364  & 3.0       & r & 364 & 3.0  \\
\rule{0pt}{2ex}ZTFJ1758+7642 & 2019-08-05 & g      & 318  & 5.0       & r & 180 & 5.0  \\
ZTFJ1758+7642 & 2019-08-07 & g      & 180  & 5.0       & r & 181 & 5.0  \\
ZTFJ1758+7642 & 2019-08-07 & g      & 453  & 5.0       & r & 453 & 5.0  \\
ZTFJ1758+7642 & 2019-10-01 & g      & 101  & 3.4       & i & 278 & 3.4 \\
ZTFJ1758+7642 & 2019-10-01 & g      & 668  & 3.4       & i & 666 & 3.4 \\
ZTFJ1758+7642 & 2020-01-22 & g      & 380  & 3.0       & r & 400 & 3.0  \\
ZTFJ1758+7642 & 2020-01-23 & g      & 342  & 6.0       & r & 343 & 6.0   \\
ZTFJ1758+7642 & 2020-01-24 & g      & 451  & 3.0       & r & 345 & 6.0   \\
ZTFJ1758+7642 & 2022-05-06 & g      & 500  & 3.0       & r & 500 & 3.0  \\
ZTFJ1758+7642 & 2022-05-06 & g      & 500  & 3.0       & r & 500 & 3.0  \\
ZTFJ1758+7642 & 2022-07-01 & g      & 737  & 3.0       &   &     &     \\
ZTFJ1758+7642 & 2022-07-01 & g      & 567  & 3.0       & r & 567 & 3.0  \\
ZTFJ1758+7642 & 2022-07-01 & g      & 600  & 3.0       & r & 600 & 3.0  \\
ZTFJ1758+7642 & 2022-07-02 & g      & 747  & 3.0       & r & 747 & 3.0  \\
ZTFJ1758+7642 & 2022-08-22 & g      & 400  & 3.0       & r & 400 & 3.0  \\
ZTFJ1758+7642 & 2022-08-22 & g      & 600  & 3.0       & r & 600 & 3.0  \\
\rule{0pt}{2ex}ZTFJ2249+0117 & 2023-07-18 & g      & 350  & 5.0       &   &     &     \\
ZTFJ2249+0117 & 2023-09-13 & g      & 157  & 5.0       & r & 158 & 5.0  \\
ZTFJ2249+0117 & 2023-09-13 & g      & 250  & 5.0       & r & 250 & 5.0  \\
\end{tabular}
\caption{All CHIMERA observations.}
\label{tab:CHIMERA}
\end{table*}

\end{appendix}

\end{document}